\documentclass[12pt,a4paper]{article}
\usepackage[utf8]{inputenc}
\usepackage[T1]{fontenc}
\usepackage{amsmath}
\usepackage{amssymb}
\usepackage{tabularx}
\usepackage{authblk}
\usepackage[title]{appendix}
\usepackage{graphicx}
\usepackage{color} 
\usepackage[linktoc=page]{hyperref}
\hypersetup{
	colorlinks=true,
	linkcolor=blue,
	bookmarks=true,
	citecolor=[rgb]{0.54,0,0}
}
\author{Yoav Zigdon}
\affil{{\normalsize \textit{Department of Applied Mathematics and Theoretical Physics, University of Cambridge, Wilberforce Road, Cambridge CB3 0WA, United Kingdom}} \\
}
\date{}

\title{A Charge Constraint in BMN}
\begin{document}
	\maketitle\begin{abstract}
		I derive a new charge constraint in  $\frac{1}{16}$-BPS sectors of the Berenstein, Maldacena, and Nastase (BMN) supersymmetric, gauged $U(N)$ matrix quantum mechanics. It was previously shown that the supersymmetric index of the model exhibits an exponential dependence on $N^2$ from the vacuum that preserves the full gauge group. I show that the reality of the associated entropy implies that the Cartan of $SO(3)$ vanishes for any $\frac{1}{16}$-BPS sector with $U(N)$ gauge symmetry.  
		The BMN charge constraint could be useful to determine whether there exists a dual supersymmetric black hole solution asymptoting to a plane-wave background.      
	\end{abstract}
	\tableofcontents
	\section{Introduction}
	In supersymmetric quantum systems, the index is the difference between the number of bosonic ground states and fermionic ground states \cite{Witten82}. This quantity is protected against quantum corrections as it is independent of coupling constants, provided the asymptotic falloff of the potential remains the same as these parameters vary. 
	
	Beyond characterizing supersymmetric gauge theories, the index promotes goals that include a) explaining the number of black hole microstates for certain black hole solutions from a dual non-gravitational theory and b) strengthening the gauge/string correspondence by successful tests, which include agreements of large-$N$ indices in holographic gauge theories, and the free energies of certain black hole solutions to gauged supergravities, that asymptote to geometry with an Anti de-Sitter (AdS) factor. 
	
	The Bogomol'nyi–Prasad–Sommerfield (BPS) ground-state entropy of the gauge theoretical systems can be extracted by utilizing a maximization procedure of a Legendre transformation of the supersymmetric index in an ensemble where charges and angular momenta are fixed~\cite{Hosseini:2017mds}. This technique gives rise to complex entropies in several examples, and common sense imposes that the imaginary part of the entropy has to vanish. This implies an equation called the ``charge constraint'' that relates the charges and angular momenta.  
	
	The first example that the entropy of a black hole was calculated from a superconformal field theory (SCFT) index is a work that preceded AdS/CFT, \cite{Strominger:1996sh}, where the large gauge group rank index of the D1-D5 SCFT was shown to match the exponential of the Bekenstein-Hawking entropy of a supersymmetric black hole, which asymptotes to a geometry containing an AdS$_3$ factor after taking a decoupling limit. 
	
	The work of \cite{Benini:2015noa} explained the Bekenstein-Hawking entropy of a black hole in asymptotically AdS$_4$ geometry from the large-$N$ topologically twisted index of the Aharony, Bergman, Jafferis, and Maldacena SCFT. Reference \cite{Cabo-Bizet:2018ehj} accounted for the Bekenstein-Hawking entropy of a $\frac{1}{16}$-BPS black hole that asymptotes to AdS$_5$ spacetime, from the index of the large-$N$, $\mathcal{N}=4$ Super-Yang-Mills (SYM) SCFT. An extension to an AdS$_7$ background appeared in \cite{Nahmgoong:2019hko} using the large-$N$ index of the $A_N$ SCFT. In all known examples, the bulk solutions satisfy a charge constraint. A fundamental explanation for this pattern has been missing; attempts to explain the validity of the charge constraint from a dual perspective appeared in \cite{Larsen:2021wnu},\cite{Larsen:2024fmp}.

	To some, it is useful to produce more examples where the number of black hole microstates admits a non-gravitational theoretic interpretation, including in gauged matrix quantum mechanics. An extrapolation of current knowledge suggests that the charges and angular momenta of the black hole solutions would satisfy a suitable constraint.  
	
	The Berenstein, Maldacena, and Nastase (BMN) matrix model is a supersymmetric quantum mechanics with the gauge group $U(N)$, global symmetry that contains $SO(3)\times SO(6)$, whose Hamiltonian includes mass and Myers terms that deform the Banks, Fischler, Shenker and Susskind (BFSS) matrix model Hamiltonian \cite{Berenstein:2002jq}. In the large-$N$ limit, the BMN model was conjectured to correspond to M-theory in an asymptotically uncompactified plane-wave background. The semiclassical limit of the model, where the deformation parameter is taken to infinity at finite $N$, divides the spectrum of the theory into superselection sectors labeled by partitions of $N$, each of which preserves a subgroup of $U(N)$ and admits an interpretation in terms of fuzzy spheres \cite{Dasgupta:2002hx}.  
	
	Three pieces of evidence in favour of the BMN conjecture above are that a) the supergvation multiplet on the M-theory side can be mapped one-to-one to the 256 center-of-mass states of the matrix theory. b) Half-BPS solutions to the low energy limit of M-theory, with $SO(6)\times SO(3)$ isometry, can be mapped to fuzzy sphere vacua or classical solutions of the gauged matrix model \cite{Maldacena:2002rb},\cite{Lin:2004nb},\cite{Lin:2005nh}. c) The existence of a non-supersymmetric black hole solution asymptoting to the plane-wave background  \cite{Costa:2014wya} is consistent with the evidence that there is a deconfined phase of the BMN model where the free energy scales with $N^2$ and the Polyakov loop is nonzero~\cite{Furuuchi:2003sy},\cite{Catterall:2010gf},\cite{Asano:2018nol} \cite{Bergner:2021goh}, \cite{Jha:2024rxz}.
	
	In 2024, reference \cite{Chang:2024lkw} utilized a calculation of the supersymmetric index of small black holes in AdS$_5$ \cite{Choi:2021lbk}, with appropriate modifications, to demonstrate that one of the contributions to the $\frac{1}{16}$-BPS index of the BMN model from the vacuum which preserves $U(N)$,  displays an exponential behavior in $N^2$. The maximization procedure explained above leads to entropy scaling also with $N^2$.  Equations (\ref{Chang}) and  (\ref{Entropy}) in Section~\ref{sec:review} present the relevant expressions for the logarithm of the index and the entropy, respectively. 
	
	Before the present note, the corresponding charge constraint in the BMN model was unknown. The main purpose of this note is to determine this constraint, motivated by finding a dual supersymmetric black hole solution that shares the same entropy. Equation (\ref{ChargeConstraint}) in Section~\ref{sec:cal} shows the main result, namely that the Cartan generator of $SO(3)$, denoted by $M^{12}$, is zero. 
	Furthermore, the indices associated with different pairs of supercharges preserved, which have not appeared in the literature, are shown to receive a universal saddle-point contribution equal to (\ref{Chang}) at large $N$. 
	
	The outline of the note is as follows. In Section \ref{sec:review}, several relevant aspects and equations of the BMN model are reviewed, and some $\mathcal{N}=4$ SYM expressions are listed. In Section \ref{sec:cal}, the BMN charge constraint is calculated when the $SO(6)$ angular momenta are equal and all angular momenta are much smaller than $N^2$. A more general calculation that follows arrives at the same result Eq.~(\ref{ChargeConstraint}). Section~\ref{sec:discussion} includes a discussion about an implication for a black hole dual.  Appendix \ref{app} includes a calculation of contributions to the indices of BPS states annihilated by different pairs of supercharges.
	
	\section{Review of Aspects of BMN}
	\label{sec:review}
	The goal of this section is to review the BMN matrix model with emphasis on its perturbative spectrum and BPS states in particular.
	
	First, the field content of the model and terms in its Hamiltonian are written.
	
	The matrix content of the BMN model is nine bosonic matrices $X^{\mu}$ and 16 fermionic matrices $\psi_{I\alpha},\psi^{\dagger I \alpha}$ where $I$ labels the fundamental representation of $SU(4)$ and $\alpha$ labels the fundamental representation of $SU(2)$.
	
	The Hamiltonian of the BMN matrix model contains, in addition to the BFSS Hamiltonian, a ``deformation term'' preserving $SO(3)\times SO(6)$ flavor symmetry, which is defined below
	\begin{equation}
		\label{def}
		H_{\mu} = \frac{R}{2} \text{Tr} \Big[ \Big( \frac{\mu}{3R}\Big)^2 \sum_{i=1} ^{3} (X^i)^2+ \Big(\frac{\mu}{6R}\Big)^2 \sum_{m=4} ^{9} (X^m)^2 + \frac{\mu}{4R} \psi^{\dagger I\alpha}  \psi_{I \alpha} + i\frac{2\mu}{3R } \epsilon_{ijk} X^i X^j X^k\Big]~.
	\end{equation}
	The parameter $R$ has the interpretation of the asymptotic radius of the dual eleven-dimensional circle, $\mu$ is the deformation parameter, $X^i$ are three of the bosonic matrices, and $X^{m}$ are the other six. The last term on the R.H.S. of Eq.~(\ref{def}) represents the Myers term and involves only $X_1, X_2$ and $X_3$. 
	
	Second, the classical solutions of the model are reviewed. In the limit where the ratio of the deformation parameter to the asymptotic radius $\frac{\mu }{R}\to \infty$ with fixed and finite $N$, the vacua of the model are \cite{Berenstein:2002jq}
	\begin{equation}
		X^m =0 ~,~ X^ i = \frac{\mu}{3R} J^i~,
	\end{equation}
	where $J^i$ form an $N$-dimensional matrix representation of $SU(2)$ and $i=1,2,3$. The representations of $SU(2)$ are generically reducible and
	each partition of $N$ amounts to a vacuum:
	\begin{equation}
		N = \sum_{k=1} ^{K} n_k N_k~,
	\end{equation}
	where the notation $k=1,..., K$ labels $K$ different irreducible representations, $n_k$ is the degeneracy of the $k$'th irreducible representation, and $N_k$ is the dimensionality of the $k$'th irreducible representation. For example, the trivial vacuum preserves the $U(N)$ gauge group is described by $K=1 ~,~ n_1 = N ~,~ N_1 =1$. In reducible vacua, the gauge group is broken to $U(n_1) \times ... \times U(n_K)~.$
	
	The perturbative spectrum of the model is reviewed next. The paper \cite{Dasgupta:2002hx} solved for the spectrum of the model in the semiclassical limit and found that bosonic and fermionic quantum harmonic oscillators generate it. About the irreducible vacuum/single-membrane vacuum characterized by $K=1, n_1=1, N_1=N$, the quantum excitations are described in Table~\ref{tab}.
	\begin{table}[h]
		\begin{center}
			\begin{tabular}{ | c |c| c | c|  } 
				\hline
				Oscillator& Mass & $SO(6)\times SO(3)$ Rep & BPS-fraction \\ 
				\hline
				$\alpha_{jm}$ & $\frac{j+1}{3}\mu$ & $(1,2j+1)$~,~$0\leq j \leq N-2$ & 0 \\ 
				\hline
				$\beta_{jm}$ & $\frac{j}{3}\mu$ & $(1,2j+1)~$,$~1\leq j \leq N$ & $\frac{1}{4}$ \\ 
				\hline 
				$x^a _{jm}$ & $\frac{j+\frac{1}{2}}{3}\mu$ & $(6,2j+1)~,~0\leq j \leq N-1$ & $\frac{1}{8}$ \\ 
				\hline 
				$\chi^{\bar{I}} _{jm}$ & $\frac{j+\frac{3}{4}}{3}\mu$ & $(\bar{4},2j+1)~,~-\frac{1}{2}\leq j \leq N-\frac{3}{2}$ & $\frac{1}{16}$ \\
				\hline 
				$\eta_{jm} ^I$ & $\frac{j+\frac{1}{4}}{3}\mu$ & $(4,2j+1)~,~\frac{1}{2}\leq j \leq N-\frac{1}{2}$ & $\frac{3}{16}$ \\
				\hline 
			\end{tabular}
		\end{center}
		\caption{Information on the spectrum of the BMN model about the irreducible vacuum. The letters $\alpha_{jm},\beta_{jm},x_{jm} ^a$ appear in expansions of the bosonic fluctuation matrices in terms of matrix spherical harmonics and create bosonic excitations, and similarly, $\eta_{jm} ^{I}$ and $\chi_{jm} ^{\bar{I}}$ generate fermionic modes.  The index $m$ runs between $-j$ to $+j$. The index $a$ labels the fundamental representation of $SO(6)$; $I$ and $\bar{I}$ label the fundamental and anti-fundamental representation of $SU(4)$, respectively.}
		\label{tab} 
	\end{table} 
	\\ Next, the supersymmetry algebra of the model is reviewed. A Cartan generator of $SO(3)$ is denoted by $M^{12}$ and the Cartan generators of $SO(6)$ are denoted by $M^{45},M^{67},M^{89}$. The model has $SU(2|4)$ supersymmetry with 32  supercharges, $16$ of which are denoted by $Q_{I \alpha}$ and $Q_{I \alpha} ^{\dagger}$ with $I=1,...,4$ and $\alpha=\pm$.
	The anti-commutations relations of these supercharges are expressible in terms of the Hamiltonian and the Cartan generators above \cite{Dasgupta:2002hx}
	\begin{align}
		\label{anticom}
		\{ Q_{I\alpha} ^{\dagger},Q_{J\beta}\} = 2\delta _{IJ} \delta _{\alpha\beta} H-\frac{\mu}{3}\epsilon^{ijk} \sigma^{k} _{\alpha\beta} \delta_{IJ} M^{ij}-\frac{i\mu}{6} \delta_{\alpha \beta} (g^{mn})  _{IJ} M^{mn}~,
	\end{align}
	where $\sigma^k$ is a Pauli matrix ($k=1,2$ or $3$) and $g^{mn}$ are Clebsch–Gordan coefficients that relate anti-symmetrized products of fundamental representations of $SU(4)$ to the fundamental representation of $SO(6)$. A simplification of formula (\ref{anticom}) is given in
	\begin{equation}
		\label{Theta}
		\Theta_{\{\epsilon_j\}}\equiv \frac{1}{2} \{ Q,Q^{\dagger}\}_{\{\epsilon_j\}}=H-\epsilon_1 \frac{\mu}{3}  M^{12} +\epsilon_2 \frac{\mu}{6}  M^{45}+\epsilon_3 \frac{\mu}{6} M^{67} - \frac{\mu}{6}\epsilon_2 \epsilon_3 M^{89}~,
	\end{equation}
	where $\epsilon_1,\epsilon_2,\epsilon_3 = \pm 1$.
	 
	The partition function and index of the model are reviewed. The grand-canonical partition function of the model is defined by
	\begin{equation}
		\label{Z}
		Z \equiv \text{Tr} \Big( e^{-\beta \Theta -2\omega M^{12} -\Delta_{45} M^{45} - \Delta_{67} M^{67} - \Delta_{89} M^{89}} \Big)~,
	\end{equation}
	where $\beta$ is an inverse temperature parameter, and $\omega,\Delta_{45},\Delta_{67},\Delta_{89}$ are the angular potentials conjugate to the angular momenta  $M^{12},M^{45},M^{67},M^{89}$, respectively. The $\frac{1}{16}$-BPS index of the theory with respect to the specific pair of supercharges with values $\epsilon_1=1,\epsilon_2=-1,\epsilon_3=-1$ chosen by \cite{Chang:2024lkw}, can be defined as $Z$ above where the trace runs only over the relevant BPS sector, subject to a constraint on the angular potentials
	\begin{equation}
		\label{Constraintchemicals}
		\Delta_{45} + \Delta_{67} + \Delta_{89} - 2\omega = 2\pi i~.
	\end{equation}
	Equation~(\ref{Constraintchemicals}) ensures that states carrying angular momenta that differ by integer multiplications of the angular momenta carried by the preserved supercharges are weighted equally in the index (up to a sign). In addition, one can multiply the R.H.S. of Eq.~(\ref{Constraintchemicals}) by any integer and still obtain a supersymmetric index.
	Substituting Eq.~(\ref{Constraintchemicals}) into Eq.~(\ref{Z}) yields the supersymmetric index when restricting the trace to be over BPS states
	\begin{equation}
		\label{index}
		\mathcal{I} = \text{Tr}_{BPS}\Big[ e^{2\pi i M^{12}-\beta \Theta-\Delta_{45} \left(M^{12}+ M^{45}\right) - \Delta_{67} \left( M^{12} + M^{67}\right)-\Delta_{89} \left( M^{12} + M^{89}\right)}\Big]~.
	\end{equation}
	One identifies the spinor number $F = 2M^{12}$, which recasts Eq.~(\ref{index}) in a more familiar form. The $\frac{1}{16}$-BPS indices of states annihilated by other supercharge pairs are considered in Appendix~\ref{app}. Below, a continuation of the review of the index of the model \cite{Chang:2024lkw} is written.\\
	The supersymmetric index admits a series representation in terms of partitions of $N$, which label the classical solutions of the model:
	\begin{equation}
		\label{IndexDec} 
		\mathcal{I}  = \sum_{\{n_i,N_i\}, \sum_{k=1} ^{K} n_k N_k=N} \mathcal{I}  _{n_i,N_i}~,
	\end{equation}
	and each contribution admits a unitary matrix integral representation
	\begin{align}
		\label{IndexInt}
		\mathcal{I} _{n_i,N_i} = \int \prod_{k=1} ^{K} [dU_k] \exp\Bigg(\sum_{m=1} ^{\infty} \sum_{k,l=1} ^{K} \frac{1}{m} z(m \Delta_{mn}) \text{Tr} \big( \big(U_{k} ^{\dagger} \big)^m \big) \text{Tr}\big(  U^m _l \big) \Bigg)~,
	\end{align}
	where the notation $U_k$ represents an $n_k \times n_k$ unitary matrix (and as stated, $n_k$ is the degeneracy of the $k$'th irreducible representation of $SU(2)$). The multi-particle index $\mathcal{I}$ is determined by the single-particle BPS index $z$ \cite{Kinney:2005ej}, which is defined in Eq.~(\ref{Z}) restricted to run over states of at most one particle. An expression for $z$ after performing the geometric series is
	\begin{align}
		\label{singleLetter}
		&z_{kl} (\omega,\Delta_{45},\Delta_{67},\Delta_{89}) =\delta_{N_k,N_l}+\frac{\Big(e^{-\omega|N_k - N_l| }-e^{-\omega(N_k+N_l)}\Big)  }{e^{2\omega}-1} \times\nonumber\\
		&  \left(1+e^{\omega + \frac{1}{2}(\Delta_{45}-\Delta_{67}-\Delta_{89})}\right) \left(1+e^{\omega + \frac{1}{2}(\Delta_{67}-\Delta_{45}-\Delta_{89})}\right)\left(1+e^{\omega + \frac{1}{2}(\Delta_{89}-\Delta_{45}-\Delta_{67})}\right)~.
	\end{align}
	In the sector which preserves $U(N)$, all $N_k~,~N_l=1$, and the constraint on the angular potentials (\ref{Constraintchemicals}) implies that the single-particle index, which is denoted by $z_{N,1}$, is given by
	\begin{align}
		\label{singleLetter1.5}
		&z_{N,1} (\Delta_{45},\Delta_{67},\Delta_{89})= 1- \big(1-e^{-\Delta_{45}}\big)\big(1-e^{-\Delta_{67}}\big)\big(1-e^{-\Delta_{89}}\big)~.
	\end{align} 
	A contribution to the supersymmetric index of BMN was calculated in \cite{Chang:2024lkw}, at large-$N$, small and fixed $\Delta_{45},\Delta_{67},\Delta_{89}$, and for the vacuum which preserves $U(N)$. The dominant saddle-point approximation is given by
	\begin{equation}
		\label{Chang} 
		\log\big( \mathcal{I} _{N,1}\big) =-\frac{3N^2}{2\pi^2} \Delta_{45} \Delta_{67} \Delta_{89}~.
	\end{equation}
	Utilizing the maximization procedure to find the entropy coming from this piece, and assuming $\frac{M^{12}+M^{mn}}{N^2} \ll1~,$ with $\{m,n\}$ being $4-5,6-7$ and $8-9$, the corresponding entropy contribution reads
	\begin{equation}
		\label{Entropy} 
		S_{N,1}  = 2\pi \sqrt{\frac{2}{3}}N^2\sqrt{\frac{M^{12}+M^{45}}{N^2}\times \frac{M^{12}+M^{67}}{N^2}\times \frac{M^{12}+M^{89}}{N^2}} ~.
	\end{equation}
	Finally, some numerical evidence was supplied that the entropy possesses an $N^2$ growth even without making the assumption $\frac{M^{12}+M^{mn}}{N^2} \ll1~.$
	\subsection{$\mathcal{N}=4$ Relevant Expressions}
	A brief review of the $\frac{1}{16}$-BPS superconformal index of $\mathcal{N}=4$ Super-Yang-Mills with $U(N)$ gauge group is written. States in this model carry three charge $Q_1,Q_2,Q_3$, whose dual chemical potentials are denoted by $\Delta_{1},\Delta_{2},\Delta_{3}$ and two angular momenta $J_1,J_2$ with conjugate angular potentials $\omega_1,\omega_2$. 
	The paper \cite{Kinney:2005ej} recast the supersymmetric index of the model in terms of a unitary matrix integral:
	\begin{align}
		& \mathcal{I}_{\text{SYM}} = \int [dU]  \exp\Bigg( \sum_{m=1} ^{\infty} \frac{1-\frac{(1- e^{-m \Delta_1}) (1-e^{-m \Delta_2}) (1-e^{-m \Delta_3})}{(1- e^{-m \omega_1}) (1-e^{- m \omega_2})}}{m} \text{Tr} ( U^m) \text{Tr} \big( ( U^{\dagger} )^{m} \big)\Bigg)~. 
	\end{align}
	In this equation, $U$ is an $N^2 \times N^2$ unitary matrix whose entries are integrated. 
	At large-$N$ and fixed chemical potentials and angular potentials, a saddle-point approximation provides the following contribution
	\begin{equation}
		\label{N=4}
		\log(\mathcal{I}_{\text{SYM,0}}) = \frac{1}{2} N^2 \frac{\Delta_{1} \Delta_{2} \Delta_{3} }{\omega_1 \omega_2}~.
	\end{equation}
	The constraint on the chemical and angular potentials is given by
	\begin{equation}
		\label{constraint} 
		\Delta_{1} + \Delta_{2} +\Delta_{3} - \omega_1 -\omega_2 = 2\pi i~.
	\end{equation}
	The maximization procedure of a Legendre transform of the index gives the entropy
	\begin{equation}
		\label{EntMax}
		S_{\text{SYM}} = \text{ext}\Big( \log(\mathcal{I}_{\text{SYM,0}}) + \sum_{j=1} ^{3} \Delta_j Q_j + \sum_{i=1} ^2 \omega_i J_i  \Big)~,
	\end{equation}
	subject to the constraint~(\ref{constraint}).
	For simplicity, consider the following parametrization of the chemical and angular potentials 
	\begin{equation}
		\label{para}
		\Delta_{1} = \Delta_{2} = \Delta_{3} = 2\beta + 2\pi i ~,~ \omega_1 = \omega_2 = 3\beta + 2\pi i ~.
	\end{equation}
	When $Q_j = Q$ and $J_i = J$, the requirement that the imaginary part of the entropy is zero, combined with the extremum condition, imposes the ``charge constraint''
	\begin{equation}
		Q^3 + \frac{N^2}{2} J^2 = \Big(\frac{N^2}{2}+3Q\Big)(3Q^2 - N^2 J)~,
	\end{equation}  
	and the entropy is
	\begin{equation}
		S(Q,J)=2\pi \sqrt{ 3Q^2 - N^2 J}~.
	\end{equation}
	For general charges and angular momenta, the same logic implies the charge constraint \cite{Cabo-Bizet:2018ehj} \footnote{Eq.~(\ref{ChargeConstraintSYM}) can be inferred from Appendix B of \cite{Cabo-Bizet:2018ehj} with their constant $\mu$ mapped to $\frac{1}{2}N^2$ and redefining their angular momenta by a minus sign.}
	\begin{align}
		\label{ChargeConstraintSYM}
		Q_1 Q_2 + Q_1 Q_3 + Q_2 Q_3-\frac{1}{2} N^2 (J_1+J_2) = \frac{ Q_1 Q_2 Q_3 +\frac{1}{2}N^2 J_1 J_2 }{Q_1+ Q_2+Q_3+\frac{1}{2}N^2}~.  
	\end{align}
	\section{Calculation}
	\label{sec:cal}
	The purpose of this section is to present a calculation of the charge constraint of the BMN sector that preserves $U(N)$ gauge symmetry and a pair of supercharges, assuming that $N$ is large. Prior to going to the derivation, the final result is given by
	\begin{align}
		\label{ChargeConstraint}
		M^{12}=0~.  
	\end{align}
	\subsection{Equal Charges and Angular Potentials}
	The simplest way to obtain the charge constraint is to set equal angular potentials in the SO$(6)$ flavour subgroup
	\begin{equation}
		\Delta_{45} = \Delta_{67} = \Delta_{89} = \Delta ~,
	\end{equation}
	and equal angular momenta conjugate to these angular potentials 
	\begin{equation}
		M^{45} = M^{67} = M^{89} = Q~.
	\end{equation}
	Introducing a Lagrange multiplier $\Lambda$, the entropy function is
	\begin{equation}
		\label{entStart} 
		S_{N,1} =  -\frac{3}{2\pi^2} N^2 \Delta^3 + 3 \Delta Q + 2\omega M^{12} + \Lambda (3\Delta - 2\omega - 2\pi i)~. 
	\end{equation}
	Substituting $\omega$ for the constraint on the angular potentials (\ref{Constraintchemicals}), the entropy expression (\ref{entStart}) becomes
	\begin{equation}
		S_{N,1} =  -\frac{3}{2\pi^2} N^2 \Delta^3 + 3 \Delta (Q+M^{12})  -2\pi i  M^{12}~.
	\end{equation}
	The maximum is obtained at 
	\begin{equation}
		\Delta_* = \frac{\pi}{N} \sqrt{\frac{2}{3}} \sqrt{Q+ M^{12}}~, 
	\end{equation}
	and the reality of the extremal entropy entails $M^{12}=0$. Its real value then reads
	\begin{align}
		S_{N,1} = 2\pi\sqrt{\frac{2}{3}} \frac{1}{N} Q^{\frac{3}{2}}~.
	\end{align}
	\subsection{Unequal Charges} 
	Next, the assumption of equal charges and angular potentials is lifted.
	At least in the case of $\mathcal{N}=4$ SYM theory, the charge constraint is obtained by starting from an expression for the supersymmetric free energy, Eq.~(\ref{N=4}), that depends on all chemical potentials, applying the maximization procedure for the entropy (\ref{EntMax}), and demanding its reality.\\ 
     Starting points are Eqs.~(\ref{Constraintchemicals}) and (\ref{Chang}) of the constraint on the chemical potentials and the piece of the supersymmetric index, which are copied here for convenience:
	\begin{equation}
		\label{Constraintchemicalsb}
		\Delta_{45} + \Delta_{67} + \Delta_{89} - 2\omega = 2\pi i~.
	\end{equation}
	\begin{equation}
		\label{Changb} 
		\log\big( \mathcal{I} _{N,1}\big) =-\frac{3N^2}{2\pi^2} \Delta_{45} \Delta_{67} \Delta_{89}~.
	\end{equation}
	 	To express the saddle-point contribution to the supersymmetric index of the BMN model in terms of all angular potentials, one can substitute each factor $\Delta_{mn}$ by an expression deduced from the constraint on the angular potentials. For example, $\Delta_{45} = - \Delta_{67}- \Delta_{89} + 2\omega + 2\pi i$, and similarly for $\Delta_{67},\Delta_{89}$. Then, the saddle-point piece for the logarithm of the supersymmetric index reads
	\begin{equation}
		\log(\mathcal{I}_{N,1}) = -\frac{3}{2\pi^2} N^2 (2\omega + 2\pi i - \Delta_{67}- \Delta_{89}) (2\omega + 2\pi i - \Delta_{45} - \Delta_{89}) (2\omega + 2\pi i - \Delta_{45} - \Delta_{67})~.
	\end{equation}
	The entropy is given by a Legendre transform of this quantity, subjected to maximization:
	\begin{align}
		\label{S} 
		S_{N,1} = \text{max}\Big(&\log(\mathcal{I}_{N,1}) + \Delta_{45} M^{45} + \Delta_{67} M^{67} + \Delta_{89} M^{89} +2\omega M^{12}\nonumber\\
		&+ \Lambda(\Delta_{45} + \Delta_{67} + \Delta_{89} -2\omega -2\pi i)\Big)~.
	\end{align}
	The last term $\Lambda(\Delta_{45} + \Delta_{67} + \Delta_{89} -2\omega -2\pi i)$ corresponds to a Lagrange multiplier enforcing the constraint on the angular potentials. The goal below is to find the extrema of the entropy, and its value at the extremum. The extremization equations are:
	\begin{align}
		\frac{\partial S_{N,1}}{ \partial \Delta_{45}} =& \frac{3}{2\pi^2} N^2 (2\omega + 2\pi i - \Delta_{67}- \Delta_{89})  (2\omega + 2\pi i - \Delta_{45} - \Delta_{67})+\nonumber\\
		&\frac{3}{2\pi^2} N^2 (2\omega + 2\pi i - \Delta_{67}- \Delta_{89}) (2\omega + 2\pi i - \Delta_{45} - \Delta_{89}) + M^{45} + \Lambda = 0~,
	\end{align}
	\begin{align}
		\frac{\partial S_{N,1}}{\partial \Delta_{67}} =& \frac{3}{2\pi^2} N^2 (2\omega + 2\pi i - \Delta_{45} - \Delta_{89}) (2\omega + 2\pi i - \Delta_{45} - \Delta_{67})\nonumber\\
		&+\frac{3}{2\pi^2} N^2 (2\omega + 2\pi i - \Delta_{67}- \Delta_{89}) (2\omega + 2\pi i - \Delta_{45} - \Delta_{89}) + M^{67} + \Lambda = 0~,
	\end{align}
	\begin{align}
		\frac{\partial S_{N,1}}{\partial \Delta_{89} } =& \frac{3}{2\pi^2} N^2  (2\omega + 2\pi i - \Delta_{45} - \Delta_{89}) (2\omega + 2\pi i - \Delta_{45} - \Delta_{67})\nonumber\\
		& +\frac{3}{2\pi^2} N^2 (2\omega + 2\pi i - \Delta_{67}- \Delta_{89})  (2\omega + 2\pi i - \Delta_{45} - \Delta_{67}) +M^{89} + \Lambda=0~.
	\end{align}
	\begin{align}
		\frac{\partial S_{N,1}}{\partial \omega} =&  -\frac{3}{\pi^2} N^2 (2\omega + 2\pi i - \Delta_{45} - \Delta_{89}) (2\omega + 2\pi i - \Delta_{45} - \Delta_{67})\nonumber\\
		& -\frac{3}{\pi^2} N^2 (2\omega + 2\pi i - \Delta_{67}- \Delta_{89})  (2\omega + 2\pi i - \Delta_{45} - \Delta_{67})\nonumber\\
		& -\frac{3}{\pi^2} N^2 (2\omega + 2\pi i - \Delta_{67}- \Delta_{89}) (2\omega + 2\pi i - \Delta_{45} - \Delta_{89})  + 2 M^{12} - 2\Lambda=0~,
	\end{align}
	\begin{equation}
		\label{constraint9}  
		\frac{\partial S_{N,1}}{\partial \Lambda} = \Delta_{45} + \Delta_{67} + \Delta_{89} - 2\omega - 2\pi i =0~.
	\end{equation}
	Plugging constraint (\ref{constraint9}) into the four equations preceding it, one obtains equations that relate the angular potentials and $\Lambda$
	\begin{align}
		\label{max1} 
		\frac{3}{2\pi^2} N^2 \Delta_{45} \big( \Delta_{67}+ \Delta_{89} \big) + M^{45} + \Lambda = 0~,
	\end{align}
	\begin{align}
		\label{max2} 
		\frac{3}{2\pi^2} N^2 \Delta_{67} \big(\Delta_{45}+ \Delta_{89}\big)   + M^{67} + \Lambda = 0~,
	\end{align}
	\begin{align}
		\label{max3} 
		\frac{3}{2\pi^2} N^2 \Delta_{89} \big( \Delta_{45} + \Delta_{67} \big)  + M^{89} +\Lambda =0~,
	\end{align}
	\begin{equation}
		-\frac{3}{\pi^2} N^2 \big( \Delta_{45} \Delta_{67} + \Delta_{45} \Delta_{89} + \Delta_{67} \Delta_{89} \big) + 2 M^{12} - 2\Lambda=0~.
	\end{equation}
	One can extract $\Lambda$ by summing over the last four equations:
	\begin{align}
		\label{Lambda}
		M^{45} + M^{67} + M^{89} + 2 M^{12} +\Lambda =0~. 
	\end{align}
	Independently, Eqs.~(\ref{max1}),(\ref{max2}) and (\ref{max3}) can be treated as three linear equations in the variables $\Delta_{45} \Delta_{67}, \Delta_{45} \Delta_{89}$ and $\Delta_{67} \Delta_{89}$. The solution of these equations is given by:
	\begin{equation}
		\label{xy}
		\Delta_{45} \Delta_{67} = -\frac{\pi^2 }{3N^2}(M^{45} +M^{67} - M^{89} + \Lambda)=\frac{2\pi^2 }{3N^2}(M^{89} + M^{12})~,
	\end{equation}
	\begin{equation}
		\label{xz}
		\Delta_{45} \Delta_{89} = -\frac{\pi^2 }{3N^2}(M^{45} + M^{89}- M^{67} + \Lambda )=\frac{2\pi^2 }{3N^2}(M^{67} + M^{12})~, 
	\end{equation}
	\begin{equation}
		\label{yz} 
		\Delta_{67} \Delta_{89} = - \frac{\pi^2 }{3N^2}( M^{67}+M^{89}-M^{45} + \Lambda)=\frac{2\pi^2 }{3N^2}(M^{45} + M^{12})~.
	\end{equation}
	The right-most sides of the last three equations come from Eq.~(\ref{Lambda}). To solve for each $\Delta_{mn}$, one can multply the last three equations to obtain
	\begin{align}
		\label{product} 
		\Delta_{45} \Delta_{67} \Delta_{89} =\pm \Big(\frac{2\pi^2}{3N^2}\Big)^{\frac{3}{2}}  \sqrt{(M^{12} + M^{45} ) (M^{12} + M^{67}) (M^{12} + M^{89})}~.
	\end{align}
	The regime of validity $\Delta_{mn} \ll 1$ implies that  $ \Delta_{45} \Delta_{67} \Delta_{89}\ll 1$, and a sufficient condition that realizes this is $M^{12} + M^{mn} \ll N^2$. 
	Keeping the plus branch of equation (\ref{product}) \footnote{The minus branch gives rise to negative entropy at the extremum, and is therefore discarded.}, one can extract each $\Delta_{mn}$ by dividing equation (\ref{product}) by Eqs.~(\ref{xy}),(\ref{xz}),(\ref{yz}). Therefore,
	\begin{equation}
		\label{Delta45} 
		\Delta_{45} = \sqrt{\frac{2}{3}} \frac{\pi}{N} \sqrt{\frac{(M^{12}+ M^{67}) (M^{12} + M^{89})}{M^{12} + M^{45}}}~.
	\end{equation}
	\begin{equation}
		\label{Delta67} 
		\Delta_{67} = \sqrt{\frac{2}{3}} \frac{\pi}{N} \sqrt{\frac{(M^{12}+ M^{45}) (M^{12} + M^{89})}{M^{12} + M^{67}}}~.
	\end{equation}
	\begin{equation}
		\label{Delta89} 
		\Delta_{89} = \sqrt{\frac{2}{3}} \frac{\pi}{N} \sqrt{\frac{(M^{12}+ M^{45}) (M^{12} + M^{67})}{M^{12} + M^{89}}}~.
	\end{equation}
	The constraint equation determines the angular potential $\omega$ to be
	\begin{align}
		\label{omega2}  
		\omega =& -\pi i + \frac{1}{2} \big(\Delta_{45} + \Delta_{67} + \Delta_{89} \big)=-\pi i + \nonumber\\
		&\frac{\pi}{N \sqrt{6}} \frac{ (M^{12}  + M^{67}) (M^{12} + M^{89})+(M^{12}+ M^{45})  (M^{12} + M^{89})+(M^{12}+ M^{45}) (M^{12}  + M^{67}) }{\sqrt{ (M^{12}+ M^{45}) (M^{12}  + M^{67}) (M^{12} + M^{89})}}~.
	\end{align}
	Let us plug the expressions for the angular potentials in terms of the angular momenta (\ref{Delta45}),(\ref{Delta67}),(\ref{Delta89}),(\ref{omega2}), into the expression for the entropy (\ref{S}):
	\begin{align}
		S_{N,1} =& - \sqrt{\frac{2}{3}} \frac{\pi}{N} \sqrt{(M^{12} + M^{45}) (M^{12}+ M^{67}) (M^{12} + M^{89})}\nonumber\\
		& +3 \times \sqrt{\frac{2}{3}} \frac{\pi}{N} \sqrt{(M^{12} + M^{45}) (M^{12}+ M^{67}) (M^{12} + M^{89})}-2\pi i M^{12}=\nonumber\\
		& = 2\pi\sqrt{\frac{2}{3}} \frac{1}{N} \sqrt{(M^{12} + M^{45}) (M^{12}+ M^{67}) (M^{12} + M^{89})}-2\pi i M^{12}~.
	\end{align}
	The requirement that the entropy is real amounts to
	\begin{equation}
		M^{12} =0~.
	\end{equation}
	\subsection{More General Entropy and Charge Constraint}
	This subsection generalizes the previous results by scanning over all possible $\frac{1}{16}$-BPS sectors characterized by different pairs of supercharges. Appendix~\ref{app} shows that the constraint on the angular potentials is generalized to
	\begin{equation}
		-\epsilon_2 \Delta_{45} - \epsilon_3 \Delta_{67} + \epsilon_2 \epsilon_3 \Delta_{89} -2\epsilon_1 \omega = 2\pi i ~.
	\end{equation}
	In this equation, $\epsilon_j=\pm 1$, which were defined in Section~\ref{sec:review} in Eq.~(\ref{Theta}), label different sectors with different pairs of supercharges.
	Additionally, by using analytical continuations, Appendix~\ref{app} illustrates that all the supersymmetric $\frac{1}{16}$-BPS indices receive the same saddle-point contribution at large-$N$, which appears in Eq.~(\ref{Chang}):
	\begin{align}
		\log(\mathcal{I}_{N,1})=-\frac{3N^2 \Delta_{45} \Delta_{67} \Delta_{89}}{2\pi^2}~.
	\end{align}
	Generalizing the computation of the previous subsection, the results for the angular potentials at the extremum are
	\begin{equation}
		\label{Delta45b} 
		\Delta_{45} = \sqrt{\frac{2}{3}} \frac{\pi}{N} \sqrt{\frac{(\epsilon_1 \epsilon_2 \epsilon_3 M^{12}+ M^{89}) (-\epsilon_1 \epsilon_3 M^{12} + M^{67})}{-\epsilon_1 \epsilon_2 M^{12} + M^{45}}}~.
	\end{equation}
	\begin{equation}
		\label{Delta67b} 
		\Delta_{67} = \sqrt{\frac{2}{3}} \frac{\pi}{N} \sqrt{\frac{(-\epsilon_1 \epsilon_2 M^{12}+ M^{45}) (\epsilon_1 \epsilon_2 \epsilon_3 M^{12} + M^{89})}{-\epsilon_1 \epsilon_3 M^{12} + M^{67}}}~.
	\end{equation}
	\begin{equation}
		\label{Delta89b} 
		\Delta_{89} = \sqrt{\frac{2}{3}} \frac{\pi}{N} \sqrt{\frac{(-\epsilon_1 \epsilon_3 M^{12}+ M^{67}) (-\epsilon_1 \epsilon_2 M^{12} + M^{45})}{\epsilon_1 \epsilon_2 \epsilon_3 M^{12} + M^{89}}}~.
	\end{equation}
	The angular potential dual to the $SO(3)$ Cartan, $\omega$, is expressible in terms of the angular potentials above through the constraint on the angular potentials:
	\begin{align}
		\omega = -\frac{1}{2} \epsilon_1 \epsilon_2 \Delta_{45} - \frac{1}{2} \epsilon_1 \epsilon_3 \Delta_{67} + \frac{1}{2} \epsilon_1 \epsilon_2 \epsilon_3 \Delta_{89} - \pi i \epsilon_1 ~.
	\end{align}
	The corresponding contribution to the (maximal) entropy for general angular momenta is
	\begin{align}
		\label{EntropyRes} 
		S_{N,1} = -2\pi i \epsilon_1 M^{12} +2\pi\sqrt{\frac{2}{3}} \frac{1}{N} \sqrt{(-\epsilon_1 \epsilon_2 M^{12} + M^{45}) (-\epsilon_1 \epsilon_3 M^{12}+ M^{67}) (\epsilon_1 \epsilon_2 \epsilon_3 M^{12} + M^{89})}~.
	\end{align}
	The end result is that for all $\frac{1}{16}$-BPS sectors preserving $U(N)$, the charge constraint reads
	\begin{equation}
		\text{Im}\big( S_{N,1}\big) =0 \Rightarrow M^{12}=0~.
	\end{equation}
	As a consequence, the entropy formula (\ref{EntropyRes}) simplifies as
	\begin{align}
		\label{EntFinal}
		S_{N,1} = 2\pi\sqrt{\frac{2}{3}} \frac{1}{N} \sqrt{ M^{45} M^{67} M^{89}}~.
	\end{align}
	Note that the regime validity of the saddle-point approximation for the index contribution is $\frac{M^{mn}}{N^2}\ll1$.
	\section{Discussion} 
	\label{sec:discussion}
	In this note, I have calculated the charge constraint Eq.~(\ref{ChargeConstraint}) in a sector of the BMN matrix model. Also, the saddle-point contribution to the large-$N$ $\frac{1}{16}$-BPS index from the $U(N)$ vacuum was shown to be independent of the $\frac{1}{16}$-BPS sector being considered. 
	
	An interesting aspect to understand is the relationship between the ``BMN sector'' of $\mathcal{N}=4$ SYM SCFT  \cite{Kim:2003rza},\cite{Choi:2023vdm},\cite{Gadde:2025yoa} and the result $M^{12}=0$ of this paper. For $Q_i\ll N^2$ and comparable $Q_i \sim Q, J_1\sim J_2\sim J$, the charge constraint of SYM implies that $J\sim Q^2$, and to zeroth order vanishes. 
	
	It is worth pointing out that the majority of the $\approx e^{2\pi  \sqrt{\frac{N}{6}}}$ contributions to the BMN supersymmetric index for large $N$ have not been calculated and, in principle, could involve cancellations which render the overall index of order one, as in the BFSS matrix model \cite{Sethi:1997pa}. Alternatively, the index is dominated by the $U(N)$ vacuum contribution. Some simplifications of Eqs.~(\ref{IndexDec}),(\ref{IndexInt}) and (\ref{singleLetter}) would potentially clarify this point.
	
	Since the charges and angular momenta of known supersymmetric black hole solutions to gauged supergravities, whose entropies were computed on the SCFT side, satisfy charge constraints, the BMN charge constraint could be beneficial for constructing such a bulk solution that preserves a pair of supercharges and is asymptotically, uncompcatified plane-wave in 11-dimensional supergravity. The Bekenstein-Hawking entropy and free energy of this putative solution are expected to agree with the logarithm of the index Eq.~(\ref{Chang}) and BMN entropy in Eq.~(\ref{EntFinal}), respectively. A few comments are in order. First, despite the lack of known examples of BPS black hole solutions that violate the relevant charge constraint, there is no theorem dictating that such configurations must be off-shell; however, the dual entropy extracted from the maximization procedure would be complex for such hypothetical examples. Second, since the $N^2$  contribution to the logarithm of the BMN index originates from the $X^m=X^i=0$ vacuum, whose spectrum of excitations is compatible with the spectrum of excitations about a single transverse M5-brane \cite{Maldacena:2002rb}, the bulk solution may share physical properties with D0-branes that form a polarized NS5-brane in Type IIA superstring theory. Third, the regime of small angular momenta may allow for establishing the existence of a solution with perturbatively small angular momenta. Fourth, a point by Jun Liu is that the near-horizon geometry of the solution may enjoy supersymmetry enhancement and preserve four supercharges, as in the case of the $\frac{1}{16}$-BPS supersymmetric black hole that asymptotes to an AdS$_5$ spacetime \cite{Gutowski:2004yv}.  It would be interesting to find out whether the black hole solution exists. 
	\subsection*{Acknowledgements}
	I thank Chi-Ming Chang for a correspondence, and Diksha Jain, Jun Liu, and Jorge Santos for their comments and discussions. This research was funded by the Blavatnik fellowship at the University of Cambridge.    
\begin{appendices} 
	\section{Various $\frac{1}{16}$-BPS Indices}
	\label{app} 
		This Appendix aims to write the contributions to supersymmetric indices from the $U(N)$ vacuum sector for eight sectors, each preserves a different pair of supercharges. 
			
	The ``Cartan vector'' packages the eigenvalues of the Hamiltonian and angular momenta for each supercharge:  
	\begin{equation}
		12\left(\frac{H}{\mu},\frac{M^{12}}{3},\frac{M^{45}}{6},\frac{M^{67}}{6},\frac{M^{89}}{6}\right)~.
	\end{equation}
	The anti-commutation relation between supercharges is encoded in $\Theta$ that was defined in Eq.~(\ref{Theta}) and will play an important role, which is copied again here:
	\begin{equation}
		\label{Theta2}
		\Theta_{\{\epsilon_j\}} = H - \epsilon_1 \frac{\mu}{3} M^{12} + \epsilon_2 \frac{\mu}{6} M^{45} + \epsilon_3 \frac{\mu}{6} M^{67} -\frac{\mu}{6}\epsilon_2 \epsilon_3 M^{89}~.
	\end{equation}
	\footnote{The sign of the second term on the R.H.S of Eq.~(\ref{Theta2}), $- \epsilon_1 \frac{\mu}{3} M^{12}$, is opposite relative to the unnumbered equation on page 26 of the paper \cite{Dasgupta:2002hx}. However, as explained in this Appendix, this sign is important to ensure the BPS nature of states in the convention that the authors chose.} In general, the Cartan vectors characterizing pairs of supercharges are
	\begin{equation}
		\label{CartanGen}
		\left(1,-2\epsilon_1,-\epsilon_2,-\epsilon_3,\epsilon_2 \epsilon_3\right)~.
	\end{equation} 
	Below, eight different cases are analyzed.\\
	
	\underline{I) $\epsilon_1=1,\epsilon_2=1,\epsilon_3=1$}:\\
	
	The supercharge Catran vector is
	$(1,-2,-1,-1,1)$. Eq.~(\ref{Theta2}) for $\Theta$ implies that in the current case, 
	\begin{equation}
		\label{Theta11}
		\Theta = H - \frac{\mu}{3} M^{12} +  \frac{\mu}{6} M^{45} +  \frac{\mu}{6} M^{67} - \frac{\mu}{6}M^{89}~.
	\end{equation}
	The set of oscillators preserving the pair of supercharges is \cite{Dasgupta:2002hx}
	\begin{align}
		&\beta~ (4j,4j,0,0,0)~,~ \nonumber\\
		&x~(4j+2,4j,-2,0,0)~,(4j+2,4j,0,-2,0)~,~(4j+2,4j,0,0,2)~,\nonumber\\
		&\chi ~ (4j+3,4j,-1,-1,1) ~,~ \nonumber\\
		&\eta~(4j+1,4j,-1,-1,-1)~,(4j+1,4j,-1,1,1)~,~(4j+1,4j,1,-1,1)~.
	\end{align}
	For example, $\beta_{jj}$ above is BPS due to the minus relative sign mentioned in Eq.~(\ref{Theta2}). One can verify that these states are all BPS by directly substituting their quantum numbers in Eq.~(\ref{Theta11}). The ranges for $j$ are:
	For $\beta_{jj}$, $1\leq j$; for $x_{jj}$, $0\leq j$; for $\chi_{jj} ^I$, $-\frac{1}{2}\leq j$; and for $\eta_{jj}$, $\frac{1}{2}\leq j$.  The grand-canonical partition function was defined in Eq.~(\ref{Z}) and allows one to compute the single-particle grand-canonical partition function $z^{+++}$ of states that preserve at least two supercharges. In the $U(N)$ vacuum sector, $z_{N,1} ^{+++}$ receives a contribution $1$ from the vacuum, and the minimal values of $j$ of all the BPS quantum harmonic oscillators contribute in excited states:
	\begin{align}
		\label{z1+++}
		z_{N,1} ^{+++} =&1+ e^{-2\omega} + e^{\Delta_{45} } + e^{\Delta_{67}} + e^{-\Delta_{89}} +e^{\omega} e^{\frac{1}{2}(\Delta_{45}+\Delta_{67} - \Delta_{89})}+\nonumber\\
		&+e^{-\omega}\left( e^{\frac{1}{2}(\Delta_{45}+\Delta_{67}+\Delta_{89})}+e^{\frac{1}{2}(\Delta_{45} -\Delta_{67} - \Delta_{89})}+e^{\frac{1}{2}(-\Delta_{45} +\Delta_{67} - \Delta_{89})}\right)~.
	\end{align}
	The constraint on the angular potentials associated with the chosen Cartan vector is
	\begin{equation}
		\label{constraint1}
		-\Delta_{45}-\Delta_{67} + \Delta_{89} - 2\omega = 2\pi i~.
	\end{equation}
	Plugging the angular potential constraint Eq.~(\ref{constraint1}) into Eq.~(\ref{z1+++}), the result for the single-particle index is
	\begin{align}
		z_{N,1} ^{+++} = 1-(1-e^{\Delta_{45}})(1-e^{\Delta_{67}})(1-e^{-\Delta_{89}})~.
	\end{align}
	From the last equation and Eq.~(\ref{IndexInt}) for the unitary integral representation of index contributions, one infers that the $U(N)$ vacuum sector contributes the following to the supersymmetric index:
	\begin{align}
		& \mathcal{I}_{N,1} ^{+++} =\int [dU] \exp\Bigg(\sum_{m=1} ^{\infty} \frac{1-(1-e^{m\Delta_{45}})(1-e^{m\Delta_{67}})(1-e^{-m\Delta_{89}})}{m}\Bigg) \text{Tr}(U^{m}) \text{Tr}(U^{\dagger m})~.
	\end{align}
	This completes case (I).\\
	
	\underline{II) $\epsilon_1=1,\epsilon_2=1,\epsilon_3=-1$:}\\
	
	The supercharge Cartan vector is $		(1,-2,-1,1,-1)$. 
	The anti-commutation relation between the supercharge and its hermitian conjugate is determined by
	\begin{equation}
		\Theta = H -\frac{\mu}{3} M^{12} + \frac{\mu}{6} M^{45} -\frac{\mu}{6} M^{67} +\frac{\mu}{6} M^{89}~.
	\end{equation}
	The BPS letters are
	\begin{align}
		&\beta~ (4j,4j,0,0,0)~,~ \nonumber \\
		&x~(4j+2,4j,-2,0,0)~,(4j+2,4j,0,2,0)~,~(4j+2,4j,0,0,-2)~, \nonumber \\ 
		&\chi ~ (4j+3,4j,-1,1,-1) ~,~ \nonumber\\
		&\eta~(4j+1,4j,-1,-1,-1)~,(4j+1,4j,-1,1,1)~,~(4j+1,4j,1,1,-1)~.
	\end{align}
	The constraint on the angular potentials is
	\begin{equation}
		\label{constraint2}
		-\Delta_{45} +\Delta_{67} -\Delta_{89} -2\omega=2\pi i ~.
	\end{equation}
	The single-letter index for the $U(N)$ vacuum sector is
	\begin{align}
		z_{N,1} ^{++-} =&1+ e^{-2\omega} + e^{\Delta_{45} } + e^{-\Delta_{67}} + e^{\Delta_{89}} +e^{\omega} e^{\frac{1}{2}(\Delta_{45}-\Delta_{67} + \Delta_{89})}+\nonumber\\
		&+e^{-\omega}\left( e^{\frac{1}{2}(\Delta_{45}+\Delta_{67}+\Delta_{89})}+e^{\frac{1}{2}(\Delta_{45} -\Delta_{67} - \Delta_{89})}+e^{\frac{1}{2}(-\Delta_{45} -\Delta_{67} + \Delta_{89})}\right)~.
	\end{align}
	The angular potential constraint Eq.~(\ref{constraint2}) gives
	\begin{align}
		z_{N,1} ^{++-} = 1-(1-e^{\Delta_{45}})(1-e^{-\Delta_{67}})(1-e^{\Delta_{89}})~.
	\end{align}
	Therefore, the $U(N)$ vacuum sector contributes the following to the supersymmetric index:
	\begin{align}
		& \mathcal{I}_{N,1} ^{++-} =\int [dU] \exp\Bigg(\sum_{m=1} ^{\infty} \frac{1-(1-e^{m\Delta_{45}})(1-e^{-m\Delta_{67}})(1-e^{m\Delta_{89}})}{m}\Bigg) \text{Tr}(U^{m}) \text{Tr}(U^{\dagger m})~.
	\end{align}
	The last equation summarizes case (II).\\
	
	\underline{III) $\epsilon_1=1,\epsilon_2=-1,\epsilon_3=1$:}\\
	
	The Cartan vector of the supercharge pair is $(1,-2,1,-1,-1)$.
	The anti-commutation relation between the supercharge and its hermitian conjugate (times a half) is
	\begin{align}
		\Theta = H -\frac{\mu}{3} M^{12} -\frac{\mu}{6} M^{45} + \frac{\mu}{6} M^{67} + \frac{\mu}{6} M^{89}~.
	\end{align}
	The BPS letters are
	\begin{align}
		&\beta~(4j,4j,0,0,0)~,~ \nonumber\\
		&x~(4j+2,4j,2,0,0)~,~(4j+2,4j,0,-2,0)~,~(4j+2,4j,0,0,-2)~, \nonumber \\
		& \chi~(4j+3,4j,1,-1,-1)~,~ \nonumber\\
		&\eta~(4j+1,4j,-1,-1,-1)~,~(4j+1,4j,1,-1,1)~,~(4j+1,4j,1,1,-1)~.
	\end{align}
	The constraint on the angular potentials is given by:
	\begin{align}
		\label{constraint3}
		\Delta_{45}-\Delta_{67} - \Delta_{89} -2\omega = 2\pi i~.
	\end{align}
	For the vacuum sector with $U(N)$ symmetry, the single-particle BPS grand-canonical partition function reads
	\begin{align}
		z_{N,1} ^{+-+} =& 1+ e^{-2\omega} +e^{-\Delta_{45}} + e^{\Delta_{67}} + e^{\Delta_{89}} +e^{\omega+\frac{\Delta_{67} + \Delta_{89}-\Delta_{45}}{2}}\nonumber\\
		&+e^{-\omega} \left( e^{\frac{\Delta_{45}+\Delta_{67} + \Delta_{89}}{2}}+e^{\frac{\Delta_{67} - \Delta_{45} -\Delta_{89}}{2}}+e^{\frac{\Delta_{89} - \Delta_{45} -\Delta_{67}}{2}}\right)~.
	\end{align}
	Substituting Eq.~(\ref{constraint3}) gives rise to a simplified single-particle index
	\begin{align}
		z_{N,1}^{+-+}=1-(1-e^{-\Delta_{45}})(1-e^{\Delta_{67}})(1-e^{\Delta_{89}})~.
	\end{align}
	As a result, the $U(N)$ vacuum sector part of the supersymmetric index is:
	\begin{align}
		& \mathcal{I}_{N,1} ^{+-+} =\int [dU] \exp\Bigg(\sum_{m=1} ^{\infty} \frac{1-(1-e^{-m\Delta_{45}})(1-e^{m\Delta_{67}})(1-e^{m\Delta_{89}})}{m}\Bigg) \text{Tr}(U^{m}) \text{Tr}(U^{\dagger m})~.
	\end{align}
	Thus, the goal of analyzing case (III) has been achieved by finding the last equation.\\
	
	\underline{IV) $\epsilon_1=1,\epsilon_2=-1,\epsilon_3=-1$:}\\
	
	The Cartan vector is $(1,-2,1,1,1)$.
	The anti-commutation relation between the supercharge and its conjugate is governed by
	\begin{equation}
		\Theta = H -\frac{\mu}{3} M^{12} -\frac{\mu}{6}M^{45}-\frac{\mu}{6}M^{67} -\frac{\mu}{6} M^{89}~.
	\end{equation}
	The angular potential  constraint is given by
	\begin{equation}
		\label{constraint4} 
		\Delta_{45} + \Delta_{67} + \Delta_{89} - 2\omega = 2\pi i~.
	\end{equation}
	The BPS letters are
	\begin{align}
		& \beta~(4j,4j,0,0,0)~,~ \nonumber\\ 
		& x~(4j+2,4j,2,0,0)~,~(4j+2,4j,0,2,0)~,~(4j+2,4j,0,0,2)~, \nonumber\\
		& \chi~(4j+3,4j,1,1,1)~,~\nonumber\\
		& \eta~(4j+1,4j,-1,1,1)~,~(4j+1,4j,1,-1,1)~,~(4j+1,4j,1,1,-1)~.
	\end{align}
	The single-letter BPS grand-canonical partition function for the $U(N)$ vacuum sector reads
	\begin{align}
		z_{N,1} ^{+--} =& 1+e^{-2\omega} + e^{-\Delta_{45}} + e^{-\Delta_{67}} + e^{-\Delta_{89}} + e^{\omega -\frac{\Delta_{45}+\Delta_{67} +\Delta_{89}}{2}} \nonumber\\ 
		&+e^{-\omega }\left(e^{\frac{\Delta_{45}-\Delta_{67} - \Delta_{89}}{2}}+e^{\frac{\Delta_{67}-\Delta_{45} - \Delta_{89}}{2}}+e^{\frac{\Delta_{89}-\Delta_{67} - \Delta_{45}}{2}}\right)~.
	\end{align}
	The imposition of the constraint  on the angular potentials Eq.~(\ref{constraint4}) gives
	\begin{align}
		z_{N,1} ^{+--}=1-(1-e^{-\Delta_{45}})(1-e^{-\Delta_{67}})(1-e^{-\Delta_{89}})~.
	\end{align}
	Hence, the contribution from the $U(N)$ vacuum sector to the supersymmetric index is
	\begin{align}
		& \mathcal{I}_{N,1} ^{+--} =\int [dU] \exp\Bigg(\sum_{m=1} ^{\infty} \frac{1-(1-e^{-m\Delta_{45}})(1-e^{-m\Delta_{67}})(1-e^{-m\Delta_{89}})}{m}\Bigg) \text{Tr}(U^{m}) \text{Tr}(U^{\dagger m})~.
	\end{align}
	The last equation is the main conclusion of the calculation in case (IV).\\
	
	\underline{V) $\epsilon_1=-1,\epsilon_2=1,\epsilon_3=1$}: \\
	
	The supercharge Catran vector is $(1,2,-1,-1,1)$. Half of the anti-commutation relation between supercharges is
	\begin{equation}
		\Theta = H + \frac{\mu}{3} M^{12} +  \frac{\mu}{6} M^{45} +  \frac{\mu}{6} M^{67} - \frac{\mu}{6}M^{89}~.
	\end{equation}
	The set of oscillators preserving the pair of supercharges is
	\begin{align}
		& \beta~ (4j,-4j,0,0,0)~,~ \nonumber \\ 
		& x~(4j+2,-4j,-2,0,0)~,(4j+2,-4j,0,-2,0)~,~(4j+2,-4j,0,0,2)~,\nonumber\\
		& \chi ~ (4j+3,-4j,-1,-1,1) ~,~ \nonumber\\
		& \eta~(4j+1,-4j,-1,-1,-1)~,(4j+1,-4j,-1,1,1)~,~(4j+1,-4j,1,-1,1)~.
	\end{align}
	The constraint on the angular potentials associated with the Cartan vector above is
	\begin{equation}
		\label{constraint5}
		-\Delta_{45}-\Delta_{67} + \Delta_{89} + 2\omega = 2\pi i~.
	\end{equation}
	The single-letter BPS grand-canonical partition function for the $U(N)$ vacuum sector is
	\begin{align}
		z_{N,1} ^{+++} =&1+ e^{2\omega} + e^{\Delta_{45} } + e^{\Delta_{67}} + e^{-\Delta_{89}} +e^{-\omega} e^{\frac{1}{2}(\Delta_{45}+\Delta_{67} - \Delta_{89})}\nonumber\\
		&+e^{\omega}\left( e^{\frac{1}{2}(\Delta_{45}+\Delta_{67}+\Delta_{89})}+e^{\frac{1}{2}(\Delta_{45} -\Delta_{67} - \Delta_{89})}+e^{\frac{1}{2}(-\Delta_{45} +\Delta_{67} - \Delta_{89})}\right)~.
	\end{align}
	Substituting the angular potential constraint Eq.~(\ref{constraint5}), the result for the single-letter index is
	\begin{align}
		z_{N,1} ^{-++} = 1-(1-e^{\Delta_{45}})(1-e^{\Delta_{67}})(1-e^{-\Delta_{89}})~.
	\end{align}
	Consequently, the $U(N)$ vacuum sector contributes the following to the supersymmetric index:
	\begin{align}
		& \mathcal{I}_{N,1} ^{-++} =\int [dU] \exp\Bigg(\sum_{m=1} ^{\infty} \frac{1-(1-e^{m\Delta_{45}})(1-e^{m\Delta_{67}})(1-e^{-m\Delta_{89}})}{m}\Bigg) \text{Tr}(U^{m}) \text{Tr}(U^{\dagger m})~.
	\end{align}
	The analysis of case (V) has been completed.\\
	
	\underline{VI) $\epsilon_1=-1,\epsilon_2=1,\epsilon_3=-1$:}\\
	
	The supercharge vector of the pair of supercharges is $(1,2,-1,1,-1)$.
	Half of the anti-commutation relation between the supercharge and its hermitian conjugate is
	\begin{equation}
		\Theta = H +\frac{\mu}{3} M^{12} + \frac{\mu}{6} M^{45} -\frac{\mu}{6} M^{67} +\frac{\mu}{6} M^{89}~.
	\end{equation}
	The BPS letters are
	\begin{align}
		& \beta~ (4j,-4j,0,0,0)~,~ \nonumber\\ 
		&x~(4j+2,-4j,-2,0,0)~,(4j+2,-4j,0,2,0)~,~(4j+2,-4j,0,0,-2)~,\nonumber\\
		&\chi ~ (4j+3,-4j,-1,1,-1)~,\nonumber\\
		&\eta~(4j+1,-4j,-1,-1,-1)~,(4j+1,-4j,-1,1,1)~,~(4j+1,-4j,1,1,-1)~.
	\end{align}
	The constraint on the angular potentials is
	\begin{equation}
		\label{constraint6}
		-\Delta_{45} +\Delta_{67} -\Delta_{89} +2\omega=2\pi i ~.
	\end{equation}
	The single-letter BPS grand-canonical partition function for the $U(N)$ vacuum sector is
	\begin{align}
		z_{N,1} ^{-+-} =&1+ e^{2\omega} + e^{\Delta_{45} } + e^{-\Delta_{67}} + e^{\Delta_{89}} +e^{-\omega} e^{\frac{1}{2}(\Delta_{45}-\Delta_{67} + \Delta_{89})}+\nonumber\\
		&+e^{\omega}\left( e^{\frac{1}{2}(\Delta_{45}+\Delta_{67}+\Delta_{89})}+e^{\frac{1}{2}(\Delta_{45} -\Delta_{67} - \Delta_{89})}+e^{\frac{1}{2}(-\Delta_{45} -\Delta_{67} + \Delta_{89})}\right)~.
	\end{align}
	The angular potential constraint Eq.~(\ref{constraint6}) implies that $z_{N,1} ^{-+-}$ is given by
	\begin{align}
		z_{N,1} ^{-+-} = 1-(1-e^{\Delta_{45}})(1-e^{-\Delta_{67}})(1-e^{\Delta_{89}})~.
	\end{align}
	Then the $U(N)$ vacuum sector contributes to the supersymmetric index:
	\begin{align}
		& \mathcal{I}_{N,1} ^{-+-} =\int [dU] \exp\Bigg(\sum_{m=1} ^{\infty} \frac{1-(1-e^{m\Delta_{45}})(1-e^{-m\Delta_{67}})(1-e^{m\Delta_{89}})}{m}\Bigg) \text{Tr}(U^{m}) \text{Tr}(U^{\dagger m})~.
	\end{align}
	This provides the desired index contribution in case (VI).\\
	
	\underline{VII) $\epsilon_1=-1,\epsilon_2=-1,\epsilon_3=1$:}\\
	
	The Cartan vector of the supercharge pair is $(1,2,1,-1,-1)$.
	The anti-commutation relation between the supercharge and its hermitian conjugate times a half is
	\begin{align}
		\Theta = H +\frac{\mu}{3} M^{12} -\frac{\mu}{6} M^{45} + \frac{\mu}{6} M^{67} + \frac{\mu}{6} M^{89}~.
	\end{align}
	The BPS letters are
	\begin{align}
		& \beta~(4j,-4j,0,0,0)~,~ \nonumber\\ 
		&x~(4j+2,-4j,2,0,0)~,~(4j+2,-4j,0,-2,0)~,~(4j+2,-4j,0,0,-2)~,\nonumber\\
		&\chi~(4j+3,-4j,1,-1,-1)~,~\nonumber\\
		&\eta~(4j+1,-4j,-1,-1,-1)~,~(4j+1,-4j,1,-1,1)~,~(4j+1,-4j,1,1,-1)~.
	\end{align}
	The constraint on the angular potentials is given by:
	\begin{align}
		\label{constraint7}
		\Delta_{45}-\Delta_{67} - \Delta_{89} +2\omega = 2\pi i~.
	\end{align}
	The single-particle BPS grand-canonical partition function for the $U(N)$ vacuum sector reads
	\begin{align}
		\label{z11--+}
		z_{N,1} ^{--+} =& 1+ e^{2\omega} +e^{-\Delta_{45}} + e^{\Delta_{67}} + e^{\Delta_{89}} +e^{-\omega+\frac{\Delta_{67} + \Delta_{89}-\Delta_{45}}{2}}\nonumber\\ 
		&+e^{\omega} \left( e^{\frac{\Delta_{45}+\Delta_{67} + \Delta_{89}}{2}}+e^{\frac{\Delta_{67} - \Delta_{45} -\Delta_{89}}{2}}+e^{\frac{\Delta_{89} - \Delta_{45} -\Delta_{67}}{2}}\right)~.
	\end{align}
	Due to Eq.~(\ref{constraint7}), Eq.~(\ref{z11--+}) reduces to
	\begin{align}
		z_{N,1}^{--+}=1-(1-e^{-\Delta_{45}})(1-e^{\Delta_{67}})(1-e^{\Delta_{89}})~.
	\end{align}
	The $U(N)$ vacuum sector part of the supersymmetric index is therefore:
	\begin{align}
		\label{I--+}
		& \mathcal{I}_{N,1} ^{--+} =\int [dU] \exp\Bigg(\sum_{m=1} ^{\infty} \frac{1-(1-e^{-m\Delta_{45}})(1-e^{m\Delta_{67}})(1-e^{m\Delta_{89}})}{m}\Bigg) \text{Tr}(U^{m}) \text{Tr}(U^{\dagger m})~.
	\end{align}
	Eq.~(\ref{I--+}) is the final step of the calculation of case (VII).\\
	
	\underline{VIII) $\epsilon_1=-1,\epsilon_2=-1,\epsilon_3=-1$}: \\
	
	The supercharge Catran vector is $(1,2,1,1,1)$~.
	The BPS states are annihilated by
	\begin{equation}
		\Theta = H + \frac{\mu}{3} M^{12} -  \frac{\mu}{6} M^{45} -  \frac{\mu}{6} M^{67} - \frac{\mu}{6}M^{89}~.
	\end{equation}
	The set of oscillators preserving the pair of supercharges is
	\begin{align}
		&\beta~ (4j,-4j,0,0,0)~,~  \nonumber\\
		& x~(4j+2,-4j,2,0,0)~,~(4j+2,-4j,0,2,0)~,~(4j+2,-4j,0,0,2)~, \nonumber\\
		& \chi ~ (4j+3,-4j,1,1,1) ~,~\nonumber\\
		& \eta~(4j+1,-4j,-1,1,1)~,(4j+1,-4j,1,-1,1)~,~(4j+1,-4j,1,1,-1)~.
	\end{align}
	The constraint on the angular potentials associated with the Cartan vector above is
	\begin{equation}
		\label{constraint8}
		\Delta_{45}+\Delta_{67} + \Delta_{89} + 2\omega = 2\pi i~.
	\end{equation}
	The single-letter BPS grand-canonical partition function for the $U(N)$ vacuum sector is
	\begin{align}
		z_{N,1} ^{---} =& 1+e^{2\omega} + e^{-\Delta_{45}} + e^{-\Delta_{67}} + e^{-\Delta_{89}} + e^{-\omega -\frac{\Delta_{45}+\Delta_{67} +\Delta_{89}}{2}} \nonumber\\
		&+e^{\omega }\left(e^{\frac{\Delta_{45}-\Delta_{67} - \Delta_{89}}{2}}+e^{\frac{\Delta_{67}-\Delta_{45} - \Delta_{89}}{2}}+e^{\frac{\Delta_{89}-\Delta_{67} - \Delta_{45}}{2}}\right)~.
	\end{align}
	The constraint Eq.~(\ref{constraint8}) brings about
	\begin{align}
		z_{N,1} ^{---}=1-(1-e^{-\Delta_{45}})(1-e^{-\Delta_{67}})(1-e^{-\Delta_{89}})~.
	\end{align}
	Hence, the contribution from the $U(N)$ vacuum sector to the supersymmetric index is
	\begin{align}
		& \mathcal{I}_{N,1} ^{---} =\int [dU] e\Bigg(\sum_{m=1} ^{\infty} \frac{1-(1-e^{-m\Delta_{45}})(1-e^{-m\Delta_{67}})(1-e^{-m\Delta_{89}})}{m}\Bigg) \text{Tr}(U^{m}) \text{Tr}(U^{\dagger m})~.
	\end{align}
	The calculation in case (VIII) is now completed.
	
	Here is a summary of the results for the $U(N)$ vacuum sector indices for all $\frac{1}{16}$-BPS sectors:
	\begin{align}
		\mathcal{I}_{N,1} ^{+++}=\mathcal{I}_{N,1} ^{-++} =\int [dU] \exp\Bigg(\sum_{m=1} ^{\infty}& \frac{1-(1-e^{m\Delta_{45}})(1-e^{m\Delta_{67}})(1-e^{-m\Delta_{89}})}{m}\Bigg) \nonumber\\
		& \times \text{Tr}(U^{m}) \text{Tr}(U^{\dagger m})~.
	\end{align}
	\begin{align}
		\mathcal{I}_{N,1} ^{++-}=\mathcal{I}_{N,1} ^{-+-} =\int [dU] \Bigg(\sum_{m=1} ^{\infty}& \frac{1-(1-e^{m\Delta_{45}})(1-e^{-m\Delta_{67}})(1-e^{m\Delta_{89}})}{m}\Bigg) \nonumber\\
		& \times \text{Tr}(U^{m}) \text{Tr}(U^{\dagger m})~.
	\end{align}
	\begin{align}
		 \mathcal{I}_{N,1} ^{+-+}=\mathcal{I}_{N,1} ^{--+} =\int [dU] \exp\Bigg(\sum_{m=1} ^{\infty}& \frac{1-(1-e^{-m\Delta_{45}})(1-e^{m\Delta_{67}})(1-e^{m\Delta_{89}})}{m}\Bigg)  \nonumber\\
		 & \times \text{Tr}(U^{m}) \text{Tr}(U^{\dagger m})~.
	\end{align}
	\begin{align}
		\mathcal{I}_{N,1} ^{+--}= \mathcal{I}_{N,1} ^{---} =\int [dU] \exp\Bigg(\sum_{m=1} ^{\infty} & \frac{1-(1-e^{-m\Delta_{45}})(1-e^{-m\Delta_{67}})(1-e^{-m\Delta_{89}})}{m}\Bigg) \nonumber\\
		& \times\text{Tr}(U^{m}) \text{Tr}(U^{\dagger m})~.
	\end{align}
	Given that the large $N$ contribution to $\log(\mathcal{I}_{N,1} ^{+--})$ contains a saddle-point piece that is proportional to $N^2\Delta_{45} \Delta_{67} \Delta_{89}$, the conclusion is that six of the other indices above can be evaluated through analytical continuations, all of which give the same contribution:
	\begin{equation}
		\label{final} 
		\log\left(\mathcal{I}_{N,1}\right)= -3N^2\frac{\Delta_{45} \Delta_{67} \Delta_{89}}{2\pi^2}~.
	\end{equation}
	Eq.~(\ref{final}) is the final result of this Appendix.
\end{appendices}

\end{document}